\documentclass[useAMS,usegraphicx]{mn2e}

\title[X-ray reflection in Ton~S180]{Investigating the reflection contribution to the X-ray emission of Ton~S180}
\author[E. Nardini, A.C. Fabian \& D.J. Walton]{E.~Nardini,$^{1}$\thanks{E-mail: enardini@cfa.harvard.edu} 
A. C. Fabian$^2$ and D. J. Walton$^2$ \\
$^1$ Harvard-Smithsonian Center for Astrophysics, 60 Garden St., Cambridge, MA 02138, USA \\
$^2$ Institute of Astronomy, Madingley Road, Cambridge CB3 0HA}

\begin{document}

\date{Released Xxxx Xxxxx XX}

\pagerange{\pageref{firstpage}--\pageref{lastpage}} \pubyear{2012}

\maketitle

\label{firstpage}

\begin{abstract}
There is now growing evidence that the soft X-ray excess is almost ubiquitous among unobscured active galaxies. 
In spite of the various interpretations that have been considered in the last years, the nature of this foremost spectral 
feature is not firmly established yet. In this context, we review from a reflection perspective the three highest-quality 
X-ray observations of the narrow-line Seyfert 1 galaxy Tonantzintla~S180, obtained by \textit{XMM-Newton} and 
\textit{Suzaku}. The X-ray spectrum of Ton~S180 shows only moderate variations over a time span of several years, 
suggesting that the same physical process accounts for the bulk of the broad-band X-ray emission at the different 
epochs, and that the properties of the X-ray source are fairly stable. We have successfully applied in our spectral 
analysis a dual-reflector model, consisting of two separate components: one arises from the surface of the accretion 
disc, is highly ionized and blurred by relativistic effects; the other is cold, quite faint, and can be associated with a 
distant reprocessor. Due to the strength and the nearly power-law shape of its soft excess emission, Ton~S180 is 
one of the most challenging sources to test the X-ray reflection scenario. In this work we provide a clear illustration 
of the great potential and spectral flexibility of blurred reflection models, while discussing some of their current limitations 
and possible shortcomings. 
\end{abstract}


\section{Introduction}

It is now widely accepted that the central engine of active galactic nuclei (AGN) consists of an accreting supermassive 
black hole (Lynden-Bell 1969; Rees 1984). In this framework, the primary X-ray emission of AGN is produced through 
the Comptonization of disc photons in a corona of relativistic electrons (Haardt \& Maraschi 1993), and can be usually 
described by means of a simple power law. With the increasing quality of the available X-ray spectra, the presence 
of other components has been progressively revealed. For instance, in sources not affected by intrinsic absorption, 
the soft X-ray emission is not invariably accounted for by just extrapolating the main power-law trend to lower energies. 
The additional component observed below 2~keV is known as soft excess (after Arnaud et al. 1985), and characterizes 
a significant fraction (possibly major; Scott et al. 2011) of X-ray unobscured AGN. The physical origin of this feature has 
to be linked to the properties of the accretion flow, but still represents a controversial issue in spite of the various models 
proposed in the recent years. \\
In terms of spectral shape, the smoothness of the soft excess is consistent with several interpretations. Direct thermal 
emission from a standard geometrically thin, optically thick accretion disc (Shakura \& Sunyaev 1973) is usually ruled 
out through a basic temperature argument, as most AGN discs should be too cool to contribute to the X-rays; yet, given 
that $T \propto (\dot{m}/M)^{1/4}$, it might become important for high accretion rates ($\dot{m}$) and relatively low black 
hole masses ($M$). This is the case for narrow-line Seyfert 1 galaxies (NLS1s; Osterbrock \& Pogge 1985), i.e. the 
sources within the type 1 population that show the lower full width at half-maximum in their optical permitted lines (FWHM 
H$\beta < 2000$ km~s$^{-1}$). As the line width distribution is continuous, the exact boundary is somewhat arbitrary; 
none the less, NLS1s represent a well-defined subclass with respect to their broad-line counterparts (BLS1s) due to 
many distinctive features (e.g. Grupe et al. 2004). In particular, the steep X-ray continuum and the prominent soft excess 
reveal a close similarity with the high/soft spectral states of black-hole binaries (Remillard \& McClintock 2006), and 
further support the speculation of nearly Eddington accretion rates in NLS1s (Pounds, Done \& Osborne 1995). On 
the other hand, the usual detection of a soft excess also in BLS1s (e.g. Porquet et al. 2004) implies that the underlying 
phenomenon is really widespread. \\
An increase of the disc effective temperature, following the Comptonization of the seed photons in a \textit{cold} 
($kT < 1$ keV), optically thick plasma, may give rise to the observed excess (e.g. Done et al. 2012). This up-scattering 
is associated with the disc photosphere itself; a tail of high-energy, non-thermal particles (Coppi 1999) can account for 
the hard power-law component as well. However, the temperature of the putative \textit{soft} corona turns out to be 
virtually constant over several orders of magnitude in black hole mass (Gierli{\'n}ski \& Done 2004). The same problem 
affects the blackbody component in the purely thermal scenario (Crummy et al. 2006). Such a universal energy hints at 
some kind of atomic process. Reflection off the surface of the accretion disc is one of the viable explanations (Fabian et 
al. 2002). In this picture, the primary X-rays go through substantial reprocessing in the photoionized upper layers of the 
disc, and emerge as a smooth continuum due to relativistic blurring in the strong gravity regime. Alternatively, it was 
suggested that the soft excess is a fake emission feature, mimicked by a broad absorption trough at $\sim$2--5 
keV ascribed to partially ionized gas along the line of sight (e.g. Middleton, Done \& Gierli{\'n}ski 2007). The detailed 
simulations of line-driven accretion disc winds are not fully consistent with the observational evidence, though (Schurch, 
Done \& Proga 2009). \\
This notwithstanding, warm absorption is very common among type 1 objects, and can hinder the characterization of 
the soft excess. Therefore, the ideal targets to investigate the nature of this component are the brightest Seyferts with 
negligible intrinsic obscuration across their entire spectrum. In this respect, the NLS1 galaxy Tonantzintla~S180 
($z=0.06198$, FWHM H$\beta$ $\simeq 900$ km~s$^{-1}$) is one of the most suitable sources. Owing to a very 
low Galactic column density ($N_\rmn{H} \simeq1.36 \times 10^{20}$~cm$^{-2}$; Kalberla et al. 2005), Ton~S180 
was the brightest AGN detected by the \textit{EUVE} satellite (Wisotzki et al. 1995; Vennes et al. 1995). This flux 
measurement, together with the steep soft X-ray spectra obtained with \textit{ROSAT} pointed observations (Fink et 
al. 1997), implied a huge emission peak in the extreme ultraviolet (UV) regime. Based on an extensive multiwavelength 
campaign, Turner et al. (2002) derived the broad-band spectral energy distribution of Ton~S180, and confirmed that the 
bulk of its energy output is released in the $\sim$10--100 eV range. If present, any X-ray warm absorber is either very 
weak or highly ionized (Turner et. al 2001; R{\'o}{\.z}a{\'n}ska et al. 2004), and is not expected to modify significantly 
the shape of the soft excess. \\
Consequently, Ton~S180 belongs to the limited sample of sources for which it is possible to probe the physical conditions 
in the very inner regions, through a comprehensive study of the intrinsic X-ray emission. Disc reflection has been shown 
to represent a basic ingredient in many of them, e.g. Mrk~335 (Larsson et al. 2008), Mrk~478 (Zoghbi, Fabian \& Gallo 
2008), Fairall~9 (Schmoll et al. 2009; Emmanoulopoulos et al. 2011), Ark~120 (Nardini et al. 2011). Indeed, the application 
of a self-consistent, dual-reflector model over the whole 0.5--40~keV energy range proved to be highly effective in our 
recent analysis of the prototypical \textit{bare} BLS1 galaxy Ark~120. Due to its striking properties and its NLS1 nature, 
hence Ton~S180 is a perfect target to further test a similar scenario. With this aim, here we review its three highest-quality 
X-ray spectra taken by \textit{XMM-Newton} and \textit{Suzaku} between 2000 and 2006. This paper is arranged as 
follows. In the next Section we summarize the observation log and data reduction. The spectral analysis is presented 
in Section 3, while Section 4 concerns the discussion of our results and their main implications. Conclusions and final 
remarks are drawn in Section 5. 

\section{Observations and data reduction}

Ton~S180 was targeted by \textit{XMM-Newton} twice, on 2000 December 14 and 2002 June 30. No significant 
background flaring occurred during the two observations, resulting in a good integration time for the EPIC pn camera 
(operated in the small-window mode) of 20.6 and 12.6~ks, respectively. A deeper monitoring by \textit{Suzaku} took 
place on 2006 December 9--12, with the source at the X-ray imaging spectrometer (XIS) nominal position; the net 
exposure was $\sim$120 ks for the XIS and 102~ks for the hard X-ray detector (HXD/PIN). The reduction of event 
files and the generation of response matrices were carried out following the standard procedures, by using the latest 
versions of the \textsc{sas} and \textsc{heasoft} packages for the \textit{XMM-Newton} and \textit{Suzaku} data 
sets. The source spectra were extracted from circular regions centred on the target, with radii of 35~arcsec for 
\textit{XMM-Newton} images and 4~arcmin for \textit{Suzaku}, while the background was evaluated over adjacent 
regions free of significant contamination. The source and background spectra from the two front-illuminated (FI) XIS 
detectors (XIS0 and XIS3) were eventually merged together. \\
Ton~S180 was marginally detected also by the HXD/PIN above 10 keV. The source count rate actually amounts 
to less than $\sim$5 per cent of the total collected events at 15--40~keV. This implies that the reproducibility of the 
instrumental non X-ray background (NXB), due to charged particles and simulated by the HXD science team with 
a nominal uncertainty of $\sim$3 per cent, may have a substantial impact on the scaling factor between the PIN 
and XIS spectra, and consequently on the spectral modelling. We have tested the NXB accuracy by extracting the 
PIN spectrum during the Earth occultation of the source. We will discuss the selection of the most appropriate NXB 
file and the implications of this choice in the following (Section~4.3). \\
The results presented in this work concern the 0.5--10~keV energy range of the EPIC pn and XIS FI spectra (from 
which the 1.75--1.95 keV interval was excluded due to calibration uncertainties), while only the 18--30~keV band has 
been conservatively taken into account for the marginal HXD/PIN detection. The MOS2 and the back-illuminated (BI) 
XIS1 spectra have been checked throughout for consistency.\footnote{In both \textit{XMM-Newton} observations the 
MOS1 detector was operated in timing mode, and it is not included in our study.} The \textsc{xspec} v12.7 fitting 
package has been employed in the spectral analysis. In order to allow the use of $\chi^2$ minimization, and 
considering that the X-ray steepness of Ton~S180 leads to non negligible noise at the higher energies, we have 
re-binned the background-subtracted spectra to a 5$\sigma$ significance in each energy channel. A minimum of 
50 counts per bin was instead adopted for the PIN data. All the uncertainties quoted below correspond to the 90 
per cent confidence range ($\Delta \chi^2 = 2.71$)  for the single parameter of interest; fluxes and related quantities 
(reflection strengths, equivalent widths) are given at the 1$\sigma$ level. 

\section{Spectral analysis}

\begin{figure}
\includegraphics[width=8.5cm]{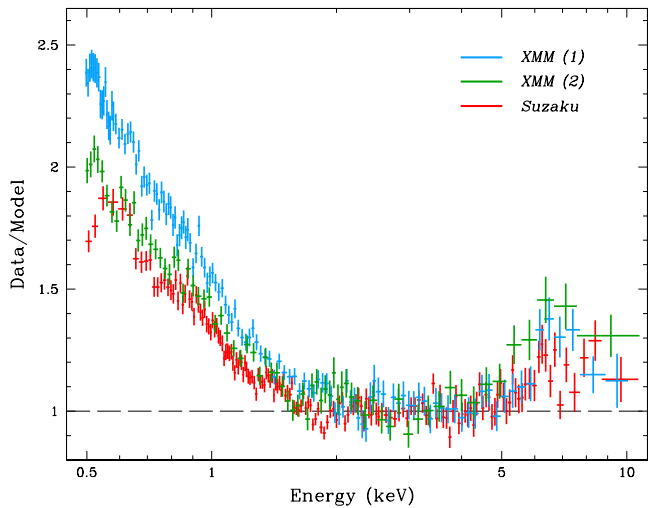}
\caption{Data/model ratios for the three spectra of Ton~S180, fitted with a power law over the 2--5~keV energy 
range: both the soft excess and a complex iron feature are clearly brought out. A common photon index is assumed, 
since the separate values ($\sim$2.33, 2.37 and 2.34) are virtually identical and consistent with each other. Though 
model-dependent, this sketch reveals the most critical points of the detailed spectral analysis, i.e. the accuracy of 
instrumental calibrations at the lower energies and the large noise above $\sim$5~keV.}
\label{rp}
\end{figure}

With the exception of the shorter \textit{XMM-Newton} exposure, the X-ray observations of Ton~S180 under review 
have already been analysed in great detail in previous dedicated works (Vaughan et al. 2002; Takahashi et al. 2010). 
These studies explored a wide range of different models to understand the nature of the broad-band X-ray emission of 
Ton~S180, and specifically of its soft excess. The strength and shape of this feature is visually indicated by the 
data/model ratio plotted in Fig.~\ref{rp}, obtained by fitting a power law over the 2--5~keV spectral range. The outcome 
is clearly model-dependent, but the low-energy upturn as well as complex iron signatures are evident. With no loss of 
accuracy, we have chosen a common hard photon index to assess the magnitude of the soft excess; its luminosity of 
$\sim$2.5--$4 \times 10^{43}$ erg~s$^{-1}$ represents at least $\sim$20--25 per cent of the observed 0.5--10~keV 
emission. To achieve a straightforward, quantitative comparison among the three spectra of Ton~S180 and their general 
properties, we have first adopted a phenomenological model consisting of a broken power law modified by neutral 
absorption, whose column density has been frozen to the Galactic value throughout this work (\texttt{tbabs} cross 
sections and solar abundances from Wilms, Allen \& McCray 2000 were assumed). The results are summarized in 
Table~\ref{t1}, and show that this baseline model provides an acceptable description of the data. Both the hard X-ray 
slope and the energy of the break are very similar in the three cases, and formally consistent within the error bars. 
A larger difference is found for the soft photon index, which follows the changes in the strength of the soft excess. 
In spite of a cumulative $\chi^2_\nu < 1.07$, clear structures in the residuals remain all across the spectra, and 
mainly in the region of iron K-shell emission and below $\sim$1.5 keV. \\
\begin{table}
\caption{Broken power-law phenomenological model. $\Gamma_\rmn{s}$: soft photon index; $\Gamma_\rmn{h}$: 
hard photon index; $E_\rmn{br}$: break energy (keV).}
\label{t1}
\begin{tabular}{l@{\hspace{30pt}}c@{\hspace{25pt}}c@{\hspace{25pt}}c}
\hline
Obs. & \textit{XMM (1)} & \textit{XMM (2)} & \textit{Suzaku} \smallskip \\
$\Gamma_\rmn{s}$ & $3.05 \pm 0.02$ & $2.94^{+0.03}_{-0.05}$ & $2.89 \pm 0.02$ \smallskip \\
$\Gamma_\rmn{h}$ & $2.28^{+0.06}_{-0.05}$ & $2.20^{+0.07}_{-0.23}$ & $2.26^{+0.02}_{-0.03} $ \smallskip \\
$E_\rmn{br}$ & $1.87^{+0.11}_{-0.13}$ & $1.84^{+0.58}_{-0.15}$ & $1.80 \pm 0.05$ \smallskip \\
$\chi^2_\nu$ & 723/665 & 512/514 & 1113/1032 \smallskip \\
\hline
\end{tabular}
\flushleft
\end{table}
Switching to a physical perspective, any of the thermal, Comptonization and reflection interpretations are plausible 
in principle for a source like Ton~S180, where absorption effects are not significant and the accretion rate is fairly 
large.\footnote{No measure of the black hole mass through reverberation mapping has been performed for Ton~S180. 
Depending on the exact value, estimated in the order of $\sim$10$^7 M_{\sun}$ (Turner et al. 2002; Satyapal et 
al. 2005) the central source radiates very close to the Eddington limit, and possibly above.} On the other hand, by 
proving that the soft excess can be fairly reproduced by nearly featureless continuum with constant slope, the 
broken power-law exercise above has two clear implications: \textit{(a)} no blackbody component with standard 
temperature\footnote{Incidentally, the best fit of the soft excess by means of a single blackbody with 
$kT \sim 0.12$--0.13~keV yields a value of $\chi^2_\nu \sim 1.17$.} would give a satisfactory description 
of the 0.5--2 keV spectral shape; \textit{(b)} any detailed Comptonization model would likely display the same 
residuals as its phenomenological counterpart, despite the useful constraints on the physical properties of the 
scattering gas. In their analysis of the first \textit{XMM-Newton} observation, Vaughan et al. (2002) included 
three blackbody components besides the power law to obtain a reasonable fit. However, the corresponding 
temperatures and luminosities dictate a linear size of the emitting region (much) lower than the innermost stable 
orbit for a Schwarzschild black hole of 10$^7 M_{\sun}$. In addition, the rapid changes of flux amplitude in the 
soft band are not compatible with any viscous process inside the disc. The same authors also explored the 
Comptonization paradigm in some depth, and it is worth reporting their findings. Any purely thermal interpretation 
requires two detached zones of different plasma temperature and optical depth to explain both the soft excess 
and the power-law continuum; this is difficult to reconcile with the observed variability pattern, which is almost 
independent of energy on short time-scales (see also Edelson et al. 2002). Allowing instead for a single Comptonizing 
region with mixed thermal/non-thermal electron distribution delivers excellent results, yet the complexity of the model is 
largely increased; no firm constraints on the nature and physical properties of this \textit{hybrid} X-ray corona can be 
derived, due to the lack of high-energy coverage. \\
Disc reflection is known to be present in Ton~S180. A broad, possibly asymmetric iron K-shell emission line has been 
detected in this source since its early observations with \textit{BeppoSAX} (Comastri et al. 1998) and \textit{ASCA} 
(Turner, George \& Nandra al. 1998). Both the skewed profile and the peak energy at $\sim$7~keV indicate an origin 
from highly ionized material close to the central source. The recent analysis of \textit{Suzaku} data suggests that the 
reflection contribution to the X-ray spectrum of Ton~S180 is presumably larger than previously thought (Takahashi et 
al. 2010). While reflection naturally accounts for the faint iron feature in the 6--7 keV range, an additional component 
was introduced in the latter study to fully model the observed emission down to $\sim$0.3 keV. In the wake of all this 
evidence, our aim is to test whether the soft excess can be interpreted as a signature of hard X-ray reprocessing as 
well. Indeed, iron fluorescence at $\sim$6.4--7~keV is just one component of the reflected spectrum. Emission lines 
from lighter elements are expected at lower energies, depending on the ionization state of the gas irradiated by the 
primary X-rays. Also, the interplay between photoelectric absorption and Compton scattering within the reflector gives 
rise to a flat, inverted continuum, which may dominate at $\sim$20--40~keV in the shape of a characteristic hump 
(e.g. George \& Fabian 1991, and references therein). All these features require a self-consistent treatment, and were 
included in our spectral analysis through the \texttt{reflionx} grid models computed by Ross \& Fabian (2005). As in 
Ark~120, we consider two different reflection components,\footnote{In the following, the reflection strength $f$ refers to 
the observed contribution to the total 0.5--10 keV emission, while the reflection fraction $R$ is computed as the intrinsic 
ratio to the power-law luminosity over the 0.1--100 keV range.} one of which is associated with the accretion disc, while 
the other is attributed to more distant matter. The former has been convolved with the \texttt{kdblur} kernel to account for 
gravitational effects close to the central black hole (according to a \texttt{laor} profile; Laor 1991). The amount of relativistic 
blurring is regulated by some key physical and geometrical properties of the disc: inner radius, inclination with respect 
to the line of sight, radial emissivity profile ($\epsilon_r \propto r^{-q}$), ionization parameter (proportional to the ratio 
between incident flux and gas density, $\xi = 4 \pi F/n_\rmn{H}$). The shape of the illuminating spectrum has been 
assumed to mirror the direct power-law continuum, and iron abundance is the same for both reflectors. \\
Our previous applications of blurred reflection models to other sources revealed the existence of considerable 
degeneracies among the critical disc parameters listed above. This holds for even relatively high-quality spectra, 
like in the case of Ark~120. In comparison, by compiling all the available \textit{XMM-Newton} and \textit{Suzaku} 
observations, the total number of counts collected from Ton~S180 is lower by roughly an order of magnitude. As a 
consequence, we expect the degeneracy issue to affect the present analysis as well, unless a model with the least 
number of variables is assumed. A recurrent argument of concern with a reflection origin of the soft excess is the high 
degree of blurring required to reproduce the remarkable smoothness of this spectral component. This may deliver a set 
of extreme parameters for the accretion disc (see the discussion below). In particular, a large inclination enhances the 
Doppler shifts along the line of sight, resulting in broader line shapes. This is actually a questionable point, as almost 
edge-on configurations are not fully conceivable for bare Seyfert galaxies, showing negligible intrinsic absorption all over 
the spectral bands. A misalignment between the plane of the disc and the large-scale absorber is possible, but in such 
a case the accretion flow itself could represent a significant source of obscuration (Lawrence \& Elvis 2010). In the light 
of these considerations, in our final model we keep the disc inclination fixed to $i = 45\degr$. Any assumption on the 
values of the other key quantities is not well justified in advance. In order to restrict the parameter space further and 
control the degeneracy problem, iron abundances in the three spectra are obviously tied together, as well as disc 
emissivity indices and inner radii. The outer border is located at 400 gravitational radii ($r_\rmn{g} = GM/c^2$), the 
maximum distance within \texttt{kdblur}, while the ionization state of the far-off reflector is frozen to $\xi_\rmn{d} = 1$ 
erg cm s$^{-1}$. Despite these demanding constraints,\footnote{We note that probability of viewing the disc with an 
inclination of $i = 45\degr$ or less is $\sim$30 per cent, as $P(i) \propto \sin i$.} we obtained an excellent fit to each 
data set, with a joint $\chi^2/\rmn{d.o.f.}=2202.3/2205$  (see Table~\ref{t2}). Even if the conditions on $\xi_\rmn{d}$ 
and $q$ are relaxed, the individual values remain fully consistent; the improvement is absolutely negligible, and does 
not compensate for the loss of degrees of freedom. In the following Section we analyse these results in the context of 
the X-ray emission of Ton~S180, and discuss their implications on the nature of the soft excess in active galaxies.
\begin{table}
\caption{Best-fitting spectral parameters of the blurred reflection model. The cumulative $\chi^2_\nu$ is 2202.3/2205. 
$\Gamma$: photon index; $q$: disc emissivity index; $r_\rmn{in}$: disc inner radius ($r_\rmn{g}$); $i$: disc inclination 
(degrees); $A_\rmn{Fe}$: iron abundance (solar units); $\xi_\rmn{b,d}$: ionization parameter of the blurred and distant 
reflectors (erg~cm~s$^{-1}$); $f_\rmn{b,d}$: 0.5--10 keV reflection strengths; $R_\rmn{b}$: 0.1--100 keV disc reflection 
fraction; $F_\rmn{obs}$: observed 0.5--10~keV flux (10$^{-11}$~erg~cm$^{-2}$~s$^{-1}$).}
\label{t2}
\begin{tabular}{l@{\hspace{25pt}}c@{\hspace{20pt}}c@{\hspace{20pt}}c}
\hline
Obs. & \textit{XMM (1)} & \textit{XMM (2)} & \textit{Suzaku} \smallskip \\
$\Gamma$ & $2.47 \pm 0.02$ & $2.36 \pm 0.03$ & $2.37^{+0.02}_{-0.01}$ \smallskip \\
$q$ & $3.85^{+0.35}_{-0.32}$(c) & $3.85^{+0.35}_{-0.32}$(c) & $3.85^{+0.35}_{-0.32}$(c) \smallskip \\
$r_\rmn{in}$ & $2.38^{+0.19}_{-0.25}$(c) & $2.38^{+0.19}_{-0.25}$(c) & $2.38^{+0.19}_{-0.25}$(c) \smallskip \\
$i$ & $45$(f) & $45$(f) & $45$(f) \smallskip \\
$A_\rmn{Fe}$ & $0.96^{+0.15}_{-0.08}$(c) & $0.96^{+0.15}_{-0.08}$(c) & $0.96^{+0.15}_{-0.08}$(c) \smallskip \\
$\log \xi_\rmn{b}$ & $2.81^{+0.06}_{-0.05}$ & $2.81 \pm 0.07$ & $2.70^{+0.03}_{-0.12}$ \smallskip \\
$f_\rmn{b}$ & $0.408 \pm 0.019$ & $0.451^{+0.029}_{-0.026}$ & $0.334^{+0.028}_{-0.012}$ \smallskip \\
$R_\rmn{b}$ & $1.09 \pm 0.11$ & $1.23^{+0.19}_{-0.17}$ & $0.77^{+0.10}_{-0.05}$ \smallskip \\
$\log \xi_\rmn{d}$ & $0.0$(f) & $0.0$(f) & $0.0$(f) \smallskip \\
$f_\rmn{d}$ & $0.039^{+0.003}_{-0.004}$ & $0.039^{+0.005}_{-0.006}$ & $0.010^{+0.002}_{-0.003}$ \smallskip \\
$F_\rmn{obs}$ & $1.53^{+0.05}_{-0.07}$ & $1.18^{+0.05}_{-0.11}$ & $1.65^{+0.08}_{-0.04}$ \smallskip \\
$\chi^2_\nu$ & $701.1/661$ & $486.4/510$ & $1014.8/1028$ \smallskip \\
\hline
\end{tabular}
\flushleft
(f): frozen value; (c): common parameter.
\end{table}

\section{Discussion}

\begin{figure}
\includegraphics[width=8.5cm]{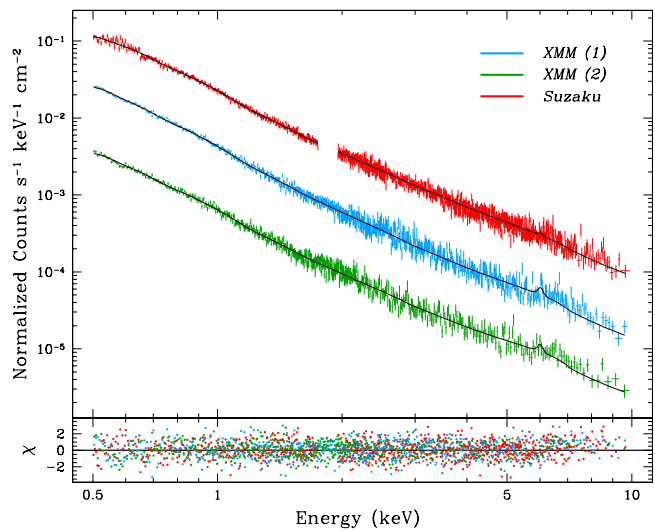}
\caption{\textit{Suzaku} XIS FI (red) and \textit{XMM-Newton} pn spectra (blue, green) of Ton~S180. As the flux levels 
during the three observations are similar, the top (bottom) spectrum has been shifted upwards (downwards) by a factor 
of five for plotting purpose. The best-fitting reflection models (black curves) are superimposed to the data, with the 
residuals (in units of $\sigma$) shown in the lower panel (the error bars have size one).}
\label{bf}
\end{figure}

It is usually very difficult to discriminate among different physical models of the soft excess on sheer statistical 
grounds, based on spectral analysis only. At the same time, any interpretation can be subject to dispute and 
eventually discarded because of some physical limitation. It is therefore important to check the reliability of a model 
in several independent ways, including its consistency with the time variability properties. Also, any fine tuning of 
the key spectral parameters (e.g. the constant disc temperature in the purely thermal scenario) should be empirically 
ruled out. Our interest in Ton~S180 is mainly driven by the fact that this NLS1 galaxy is one of the most challenging 
sources to test the blurred reflection hypothesis. The three 0.5--10~keV spectra are shown in Fig.~\ref{bf} with their 
best-fitting models. We have selected a lower limit of 0.5 keV in our analysis to avoid possible uncertainties in the 
calibration of the detectors near the bottom edge of their operational range. Irrespectively of the model adopted, in 
fact, the structures in the residuals become quite different below $\sim$0.7 keV, for both the \textit{XMM-Newton} 
MOS2/pn spectra and the \textit{Suzaku} BI/FI ones. This is clearly a sensitive point, as the presence of any 
calibration systematics forces the reflection components to reproduce spurious features in the soft excess. In any 
case, we have checked for completeness the spectral trend down to 0.3 keV. The extrapolation of our best-fitting 
model fails to account for the entire observed emission, since the excess further extends at low energies with some 
hints of flattening but no apparent turnover. An adjustment of the current parameters or a larger reflection complexity 
can be invoked, but a low-temperature thermal contribution is also likely. A blackbody with $kT \sim 16$ eV was 
needed in some of the soft X-ray spectral states during the \textit{ROSAT} campaign, and could represent the tail 
of the extreme UV emission, as the doubling time of amplitude variations is very similar in the two bands (Fink et al. 
1997; Hwang \& Bowyer 1997). For the sake of discussion, in the following we consider our model as a genuine 
upper limit to X-ray reflection in Ton~S180. \\
In general, the broad-band values of the blurring parameters are quite unusual if compared with the application of 
the same relativistic smearing to broad iron lines alone. In the latter case, both the disc inner radius and the emissivity 
index are indicative of a mild gravitational regime, with no requirement of $r_\rmn{in} < 6\ r_\rmn{g}$ or $q > 3$ (e.g. 
Nandra et al. 2007). With respect to these standards, blurred reflection models do indeed deliver \textit{extreme} 
parameters (Walton, Reis \& Fabian 2010). This is a major matter of debate, since  inner radii significantly smaller 
than the innermost stable orbit in a Schwarzschild metric hint at nearly maximally rotating black holes. We also note 
that $q = 3$ is the classical limit at large distance for a point-like, on-axis illuminating source; a flatter response is then 
expected for either a great lamppost height ($h \gg r$) or an extended/patchy corona (possibly as a result of several 
hotspots or magnetic flares). On the contrary, steep emissivity profiles imply that most of the energy is released in the 
very inner regions, within a few gravitational radii from the central engine. In addition, the reflection fraction ($R_\rmn{b}$) 
can be high, and occasionally the direct power-law continuum is completely suppressed (Fabian et al. 2012). Most of 
these puzzling aspects are actually self-consistent, and have a reasonable physical explanation in the context of 
strong light bending (Miniutti \& Fabian 2004; Wilkins \& Fabian 2011). In fact, the illuminating radiation can be diverted 
towards the inner disc below more or less effectively, depending on the vertical height of the primary source. Any 
correlation among these quantities ($r_\rmn{in}$, $q$, $R_\rmn{b}$) would represent a substantial confirmation of 
this light bending/blurred reflection scenario. The extensive variability studies conducted on Ton~S180 suggest that 
the bulk of X-ray emission occurs within the inner $\sim$20--30 $r_\rmn{g}$ (Romano et al. 2002; Edelson et al. 2002), 
and strongly support the picture of a very compact source. \\
It is important, however, to keep in mind that the computation of the reflection grids is based itself on several 
assumptions, which necessarily oversimplify the physical and geometrical properties of the accretion flow (see 
Ross \& Fabian 2005 for details). A plane-parallel slab with constant density $n_\rmn{H} = 10^{15}$ cm$^{-3}$ 
is a good approximation for the outer surface of an accretion disc, yet different values of $n_\rmn{H}$ and 
inhomogeneities are possible in sources accreting at nearly Eddington rates like Ton~S180. The same properties 
are less adequate for the cold, distant reflector, which has a minor but non negligible contribution, especially at the 
lower energies (where it may account for small departures from a perfectly smooth shape). In this case, a clumpy 
structure with lower density and optical depth would be definitely more reasonable. A further limitation that mainly 
affects the soft band is the lack of leverage on atomic abundances, which are frozen to solar values for all the 
elements other than iron. The overall shape of the reflection spectrum also depends on crucial ingredients that 
remain basically unknown, such as the role of magnetic fields, the geometry of the illuminating source, the exact 
nature of the interactions between the disc and the corona. \\
In this regard, once the goodness of the working approximations has been established, the spectral analysis can 
provide valuable information to understand the mechanisms of energy production and the properties of the X-ray 
emitting region. We have therefore explored more complex configurations, introducing in turn a broken power-law 
emissivity profile for the disc, a radial dependence then local inhomogeneities of its ionization state, gradients of 
iron abundance, but we did not obtain any stringent constraint. Finally, we have also attempted at thawing the 
disc inclination. The resulting value is rather large, consistent with an edge-on geometry, while the overall statistical 
improvement is formally significant but still limited. This is an obvious effect of the complex degeneracy among the 
parameters, as the other key quantities change in an unpredictable manner. Specifically, the disc inner radius drops 
to the innermost stable circular orbit of a Kerr black hole, and the emissivity profile is now quite flat ($q \simeq 2$), 
more suitable for an extended corona. Iron abundance becomes mildly subsolar ($A_\rmn{Fe} \simeq 0.7$), in contrast 
with both the typical values found among NLS1 galaxies and the intense Fe~\textsc{ii} optical emission of Ton~S180 
itself (Comastri et al. 1998). Hence, a model with no geometrical constraint delivers a full set of unusual parameters, 
and can be ultimately rejected on physical grounds. If the large inclination is real, instead, a possible explanation 
involves the dragging of inertial frames, for which the reflected photons tend to escape preferentially along the plane 
of the disc. In this view, all the other properties being equal, nearly edge-on sources should have a more prominent 
soft excess. It is worth noting that the discrepancy is not related with the statistical goodness around the energy of 
the broad iron line (see below). The model with free $i$ actually provides an inferior fit to the data up to $\sim$7.8 
keV in the rest frame. A marginally better match is recovered only beyond, where the spectra are somewhat flatter, 
but the noise is fairly large and the background contamination is likely significant. \\
Summarizing, the moderate spectral variations suggest that a single (or at least dominant) physical process is at 
work in the centre of Ton~S180; our basic dual-reflector model gives an effective description of its X-ray emission over 
different epochs, without requiring any \textit{extreme} condition in spite of the underlying assumptions and limitations. 
The spectral evolution follows the changes of the primary photon index ($\Delta \Gamma \simeq 0.1$) and those of 
the disc ionization, whose extent ($\sim$30--40 per cent) is in perfect agreement with the range of fluctuations observed 
for the average 0.5--10~keV flux. This kind of feedback may also account for the secular trend revealed in the softness 
ratios (Romano et al. 2002). Below we focus on the most critical points of our analysis, which illustrate some possible 
shortcomings of the blurred reflection picture. 

\subsection{Black hole spin and broad iron feature}

By assuming that the disc extends down to the innermost stable circular orbit and that no emission comes from the 
plunging region (e.g. Reynolds \& Fabian 2008), an inner border of $\sim$2.4~$r_\rmn{g}$ is not consistent with a 
Schwarzschild (non-rotating) black hole. This should not be surprising, even if there are still large uncertainties on the 
theoretical spin distribution based on the models of formation and evolution of supermassive black holes through 
mergers and/or accretion episodes (Berti \& Volonteri 2008; King, Pringle \& Hofmann 2008). Moreover, a putative 
disc truncation farther away from the centre is only associated with very low Eddington rates, while evaporation is 
ruled out by the small disc temperature. In order to assess the black hole spin directly, we switched to \texttt{kerrconv} 
(Brenneman \& Reynolds 2006), an alternative relativistic kernel that allows the dimensionless spin parameter 
$a=cJ/GM^2$ to vary ($J$ and $M$ are the black hole angular momentum and mass). According to this choice, 
we obtained that $a = 0.85^{+0.07}_{-0.05}$, but a maximally rotating black hole cannot be rejected at the 99 
per cent confidence level ($a > 0.75$). This must be taken just as a coarse estimate: even without considering 
any interpretation bias (Walton et al. 2012), the systematic errors on the present measurements of black hole spin, 
related to the modelling of the accretion and reflection physics, are thought to exceed by far the statistical ones. Our 
assumptions (e.g. $i=45\degr$) and any residual degeneracy of the parameters represent a supplementary source 
of uncertainty. It is therefore worth examining the response of our model to variations of the inner radius. We have 
frozen $r_\rmn{in} = 10\ r_\rmn{g}$, while dropping the previous constraint on the disc inclination. This test results in 
a fully acceptable fit ($\chi^2_\nu < 1.01$), and the corresponding set of key variables is now $q \simeq 5$, 
$A_\rmn{Fe} \simeq 1.2$, $\log \xi_\rmn{b} \sim 2.6$--2.8 and $i \simeq 35\degr$. The steep emissivity is physically  
inconsistent with the imposed truncation, but this is a further caveat against the face values reported in Table~\ref{t2}, 
since the parameter space has many equivalent local minima and the strict application of the $\chi^2$ criterion can 
be misleading. \\
\begin{table}
\caption{Fit of the broad iron line over the 3--8~keV range with a \texttt{laor} profile ($q=3$, $i=45\degr$; 
see Table~\ref{t2}). $E_\rmn{L}$: line energy (keV); EW$_\rmn{L}$: equivalent width (eV).}
\label{t3}
\begin{tabular}{l@{\hspace{30pt}}c@{\hspace{25pt}}c@{\hspace{25pt}}c}
\hline
Obs. & \textit{XMM (1)} & \textit{XMM (2)} & \textit{Suzaku} \smallskip \\
$\Gamma$ & $2.16^{+0.12}_{-0.08}$ & $1.97^{+0.15}_{-0.11}$ & $2.22^{+0.07}_{-0.05}$ \smallskip \\
$r_\rmn{in}$ & $64^{+47}_{-30}$(c) & $64^{+47}_{-30}$(c) & $64^{+47}_{-30}$(c) \smallskip \\
$E_\rmn{L}$ & $6.86^{+0.13}_{-0.11}$ & $6.71^{+0.40}_{-0.16}$ & $6.75^{+0.13}_{-0.28}$ \smallskip \\
EW$_\rmn{L}$ & $228 \pm 129$ & $207^{+154}_{-152}$ & $108 \pm 67$ \smallskip \\
$\chi^2_\nu$ & $179.1/199$ & $100.3/110$ & $425.9/425$ \smallskip \\
\hline
\end{tabular}
\flushleft
\end{table}
Closely related to the measure of the black hole spin is the presence of a broad iron emission feature, which has 
been regularly detected in all the X-ray observations of Ton~S180. Within a pure reflection frame, both the soft 
excess and the broad component in the iron K$\alpha$ line profile arise from the innermost regions of the accretion 
disc. In a broad-band spectral analysis, the best-fitting blurring and ionization parameters are mainly driven by the 
much larger statistical weight of the soft excess. A general agreement is expected when the iron feature is fitted 
separately, though. As mentioned above, this is not always the case, and could represent one of the most serious 
issues with the reflection scenario. Some residual structure in the $\sim$6--7~keV region is apparent in Ton~S180 as 
well, especially in the two \textit{XMM-Newton} observations (Fig.~\ref{bf}). We then focused on the 3--8 keV spectral 
range, and applied a simple model consisting of a power law plus a \texttt{laor} profile. This allows us to put tighter, 
independent constraints on the significance of the broad iron line, even if some of its properties have to be assumed. 
The results are summarized in Table~\ref{t3}. By considering its equivalent width, the line is detected at the 1.7$\sigma$ 
confidence level at most. This is consistent with all the previous observational records, included those obtained by 
\textit{ASCA} and \textit{BeppoSAX}. The profile is very broad but it is difficult to appreciate any asymmetry due 
to the moderate prominence. In both the \textit{XMM-Newton} spectra an equivalent fit can be recovered by replacing 
the disc line with a Gaussian of width $\sigma \sim 0.4$. In a Keplerian frame, this entails a FWHM broadening of 
$\sim$4~$\times 10^4$ km s$^{-1}$, corresponding to a distance of $\sim$50--60~$r_\rmn{g}$ from a black hole of 
10$^7 M_{\sun}$. Interestingly, this is the location of the reflecting clouds that have been recently proposed as the 
source of the UV emission peak in AGN (Lawrence 2011). \\
In conclusion, the broad iron line in Ton~S180 is not sufficiently strong to prove its consistency with the soft excess, 
probably because of the high ionization of the reflecting gas. The same argument is invoked to explain the lack/faintness 
of narrow iron lines, whose detection would likely cancel out the observed residuals. This hints at a range of ionization 
states across the different spatial scales much more complex than the one simulated in our analysis. None the less, 
the tentative discrepancy between the two main disc features remains a pivotal issue of blurred reflection models, and 
will be addressed in detail in a future work. 
\begin{figure}
\includegraphics[width=8.5cm]{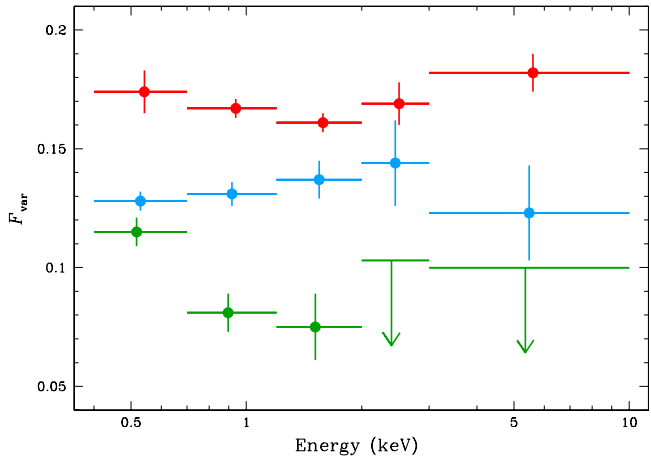}
\caption{Fractional rms variability during the three observations in the 0.4--0.7, 0.7--1.2, 1.2--2, 
2--3 and 3--10~keV energy bands. From top to bottom: \textit{Suzaku}, \textit{XMM-Newton} (1) 
and \textit{XMM-Newton} (2). The adopted time resolution is 500 s.}
\label{tv}
\end{figure}

\subsection{X-ray time variability}

The analysis of time variability and its dependence with energy is a powerful, alternative tool either to confirm or 
to discard any physical interpretation based on spectral modelling. In very favourable conditions, this complementary 
approach can even be conclusive in distinguishing between reflection and Comptonization scenarios. This possible 
discrimination relies upon the detection of time lags among the different X-ray energy bands. In particular, a delay 
between the hard X-ray variations and the soft ones at a given time-scale is regarded as the typical reverberation 
signature, corresponding to the light travel time from the illuminating source to the reflector and back. Evidence for 
\textit{negative} (soft) lags has rapidly grown in recent times (Fabian et al. 2009; Zoghbi et al. 2010; De Marco et al. 
2012), but none was found in Ton~S180 down to the limit of 500 s (Vaughan et al. 2002). \\
The timing properties of Ton~S180 have already been explored in the dedicated studies mentioned above. 
In Fig.~\ref{tv} we reproduce the fractional rms variability amplitude ($F_\rmn{var}$) of the source: this is an 
indicator of the intrinsic \textit{noise} after the measurement uncertainties have been subtracted (e.g. Edelson 
et al. 2002; Vaughan et al. 2003). For ease of comparison among the three observations here we have defined only 
five energy bands, but our findings are in full agreement with previous works where a finer sampling was adopted. 
Although the absolute values of $F_\rmn{var}$ are different, roughly scaling in proportion to the length of the single 
observations,\footnote{The power spectral density decreases with frequency as $P(\nu) \propto \nu^{-\alpha}$, with 
$\alpha \sim 1$--2 (e.g. Lawrence \& Papadakis 1993), hence the source variability is mostly sensitive to the longer 
time-scales.} the variability patterns are reasonably well matched. While there is no appreciable trend with energy for 
the two deeper data sets, implying that the spectrum varies in first approximation as a whole, the shorter 
\textit{XMM-Newton} monitoring shows a provisional decline. Rather than physical, this is presumably a 
statistical/systematic effect. Because of the spectral steepness, the count rate gets rapidly smaller and the weight 
of Poisson noise in the hard bands is considerably larger. This might result in a decrease of $F_\rmn{var}$, highly 
accentuated by both the short exposure and the low flux state characterizing this observation, which lead to the 
detection of only marginally positive excess variance above 2 keV. On the other hand, a genuine drop of $F_\rmn{var}$ 
with energy would be apparently in contrast with a reflection context. In fact, if the soft emission is a form of 
reprocessing of hard X-rays, the lower energies should reveal somewhat smoothed variations. Such behaviour is 
expected whenever the emission in a definite band can be plainly associated with a dominant reflection component, 
and also applies to the iron region (e.g. Vaughan \& Fabian 2004) and the higher energies, especially in the presence 
of a prominent Compton hump (which is actually not the case for Ton~S180, see below). \\
In general, however, the dependence of $F_\rmn{var}$ with energy does not necessarily have any intuitive interpretation. 
Both the direct emission and its blurred reflection counterpart yield a broad-band contribution, which gives rise to the 
typical soft excess plus hard power law shape (see Fig.~\ref{mc}). Under extreme gravity conditions, their interplay may 
have quite complicated manifestations in the time domain. This can be easily verified by rendering explicit the two 
components within the definition of excess variance itself, and keeping in mind that the X-ray illumination experienced 
by the disc is much more intense than the one received at infinity. As for Ton~S180, where severe light bending does not 
seem to be a decisive ingredient, the continuum irradiating the disc is almost identical to the observed, direct power law. 
The absence of time delays is still consistent with a source height of several gravitational radii, tightly corresponding to 
the values inferred from the spectral analysis for $r_\rmn{in}$, $q$ and $R_\rmn{b}$. The two perspectives are then 
compatible with the transitional regime identified by Miniutti \& Fabian (2004), hinting at a stationary source of varying 
luminosity. A distance above the disc $h \gg 10~r_\rmn{g}$ (with a nearly constant reflection opposed to a markedly 
variable power law) would instead deliver clearer structures in $F_\rmn{var}$ as well as detectable spectral changes. This 
minimal scatter in the variability pattern (see Ponti et al. 2011) has a couple of remarkable implications. First, it justifies a 
time-averaged spectral analysis like the one performed in this work, as this is fairly representative of the physical state of 
Ton~S180 during each observation. Second, it involves a tight causal connection between the soft and hard X-ray emission, 
which again fits into a reflection scenario, even if the details are partially obscure. Also some kind of complex, non-thermal  Comptonization process may be able to account for these properties in a self-consistent way, so we argue that this other 
option cannot be completely ruled out. 

\subsection{\textit{Suzaku} HXD/PIN detection}

\begin{figure}
\includegraphics[width=8.5cm]{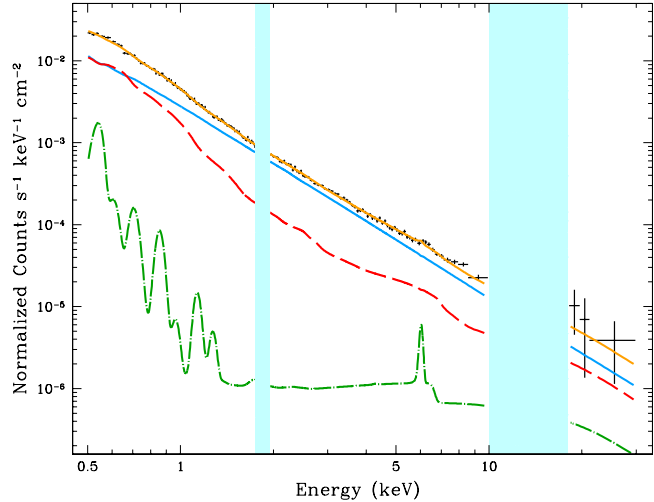}
\caption{\textit{Suzaku} XIS FI and HXD/PIN spectra (rebinned for plotting purposes) of Ton~S180. The PIN 
background was estimated during the Earth occultation of the target. The resulting high-energy spectrum is fully 
consistent with the extrapolation of low-energy best-fitting model (orange curve). The relative contribution of the 
different components is disentangled: direct power law (blue, solid), blurred reflection (red, dashed) and distant 
reflection (green, dot-dashed).}
\label{mc}
\end{figure}

As mentioned earlier, the detection of Ton~S180 beyond 10~keV is quite critical. Since the HXD/PIN is a collimating 
instrument with no imaging capabilities, the background level cannot be derived from the data records themselves. 
While the cosmic contribution has been simulated according to standard spectral forms (e.g. Boldt 1987), a tuned 
model of the particle non X-ray background (NXB) was provided by the HXD science team. Whenever the target is 
very faint, as for Ton~S180, the PIN count rate is dominated by the NXB. The accuracy of its reproducibility (typically 
within $\sim$3--5 per cent; Kokubun et al. 2007) is therefore a major factor of uncertainty, which mainly affects the 
absolute normalization of the high-energy emission of the source. In the present case, the PIN statistical weight is 
virtually negligible, and the best fit is driven by the XIS spectrum regardless of the model. As a result, we found a 
systematic PIN offset. By leaving it free to vary, we obtain a PIN/XIS cross calibration of $\sim$2.6($\pm 0.7$), 
against a recommended value of 1.16. Such disparity can be actually explained once the NXB uncertainty is taken 
into account. This point has been already addressed in great detail by Takahashi et al. (2010), who eventually 
introduced a Compton-thick partial-covering absorber to retain the nominal NXB. The latter solution is not strictly 
required, though. Moreover, it implies a considerable fine tuning in terms of both column density and covering 
fraction to reproduce the high-energy emission without modifying the spectral shape below 10 keV. The presence 
of this heavy absorption component is not envisaged in a \textit{bare} Seyfert galaxy. Yet the putative hard excess 
cannot be interpreted like a poorly modelled reflection hump, as suggested for NGC~1365 (Risaliti et al. 2009; 
Walton et al. 2010). \\ 
We did not pursue this issue further. Instead, we extracted the Earth-occulted PIN spectrum to test the NXB 
accuracy. This is an observational measure of the instrumental background, and was employed throughout 
our analysis. The outcome is shown in Fig.~\ref{mc}. There is still some discrepancy, but now the best scaling 
factor would be $\sim$1.9($\pm 1.1$), broadly consistent with the usual values; the probability of chance 
improvement following the loss of one d.o.f. is as high as 25 per cent. The existence of a hard excess in 
Ton~S180 remains somewhat controversial then, and can only be revealed by forthcoming X-ray missions 
as \textit{NuSTAR} and \textit{Astro-H}. 

\section{Conclusions}

We have presented the joint spectral analysis of the three highest-quality X-ray observations of the NLS1 galaxy 
Ton~S180, obtained by \textit{XMM-Newton} and \textit{Suzaku}. This is one of the brightest sources with negligible 
intrinsic absorption, and represents an ideal target to investigate the nature of a striking X-ray spectral feature like the 
soft excess. Even though the observations under review cover a period of several years, the 0.5--10~keV emission 
of Ton~S180 does not exhibit dramatic variations. In first approximation, the spectral shape can be described by a 
broken power law. The hard photon index ($\Gamma_\rmn{h} \simeq 2.2$--2.3) is almost constant, with an ordinary 
value among NLS1s; the soft one is steeper ($\Gamma_\rmn{h} \sim 3$) and more variable, following the strength 
of the soft excess below $\sim$2 keV. A broad iron fluorescence line hints at reflection from highly ionized material 
close to the central black hole. However, this feature is not very prominent, and falls in a noisy region owing to the 
steepness of the primary X-ray continuum. Conversely, the soft excess emission is remarkable in both strength and 
shape, challenging a possible origin from the inner disc as a blurred reflection signature. In order to probe the soundness 
of this scenario, we have applied a self-consistent model that provides a successful description of all the different data 
sets by allowing for a dual-reflector geometry. The high-ionization component ($\log \xi_\rmn{b} \simeq 2.7$--2.8) is 
associated with the disc, and accounts for most of the soft excess. Also involved is a much fainter, neutral component, 
arising from the reprocessing of the primary radiation at much larger distance from the X-ray source. The accretion flow 
extends down to $r_\rmn{in} \simeq 2.4\ r_\rmn{g}$, and this is a compelling evidence in favour of a rapidly rotating 
black hole. The substantial degeneracy among the key parameters prevents an accurate determination of the spin 
and a complete characterization of the central engine, none the less the overall picture is clearly established. \\
This work, together with our previous study of the BLS1 galaxy Ark~120, confirms that reflection models are 
extremely effective in reproducing a wide range of X-ray spectral properties. A uniform analysis of a sizable 
sample of similar objects is anyway mandatory to uncover the expected correlations (e.g. between soft excess 
and broad iron feature, reflection fraction and black hole spin) and rule out any fine tuning (e.g. of the disc ionization 
state and/or inclination). Even this kind of effort might not be conclusive if the soft excess is a complex, non-linear 
combination of all the physical processes that take place in the central regions of active galaxies. In the absence 
of a universal explanation, only the next-generation X-ray satellites would have the capabilities to disentangle the 
different contributions from thermal emission, reflection, scattering and (possibly) absorption. 

\section*{Acknowledgments}

The authors would like to thank the anonymous referee for providing helpful comments. 
EN acknowledges the financial support from NASA grants NNX10AF50G and GO0-11017X. 
ACF thanks the Royal Society. DJW acknowledges the financial support provided by the STFC.



\label{lastpage}


\begin{thebibliography}{}
\bibitem[\protect\citeauthoryear{Arnaud et al.}{1985}]{1985MNRAS.217..105A} 
Arnaud K.~A., et al., 1985, MNRAS, 217, 105
\bibitem[\protect\citeauthoryear{Berti 
\& Volonteri}{2008}]{2008ApJ...684..822B} Berti E., Volonteri M., 2008, ApJ, 684, 822
\bibitem[\protect\citeauthoryear{Boldt}{1987}]{1987PhR...146..215B} Boldt 
E., 1987, PhR, 146, 215
\bibitem[\protect\citeauthoryear{Brenneman 
\& Reynolds}{2006}]{2006ApJ...652.1028B} Brenneman L.~W., Reynolds C.~S., 2006, ApJ, 652, 1028
\bibitem[\protect\citeauthoryear{Comastri et 
al.}{1998}]{1998A&A...333...31C} Comastri A., et al., 1998, A\&A, 333, 31
\bibitem[\protect\citeauthoryear{Coppi}{1999}]{1999ASPC..161..375C} Coppi P. S., 1999, in Poutanen J., Svensson R., eds, ASP 
Conf. Ser. Vol. 161, High Energy Processes in Accreting Black Holes. Astron. Soc. Pac., San Francisco, p. 375
\bibitem[\protect\citeauthoryear{Crummy et al.}{2006}]{2006MNRAS.365.1067C} 
Crummy J., Fabian A.~C., Gallo L., Ross R.~R., 2006, MNRAS, 365, 1067
\bibitem[\protect\citeauthoryear{De Marco et 
al.}{2011}]{2012arXiv1201.0196D} De Marco B., Ponti G., Cappi M., Dadina 
M., Uttley P., Cackett E.~M., Fabian A.~C., Miniutti G., 2011, arXiv, 
arXiv:1201.0196
\bibitem[\protect\citeauthoryear{Done et al.}{2012}]{2012MNRAS.420.1848D} 
Done C., Davis S.~W., Jin C., Blaes O., Ward M., 2012, MNRAS, 420, 1848 
\bibitem[\protect\citeauthoryear{Edelson et 
al.}{2002}]{2002ApJ...568..610E} Edelson R., Turner T.~J., Pounds K., 
Vaughan S., Markowitz A., Marshall H., Dobbie P., Warwick R., 2002, ApJ, 
568, 610
\bibitem[\protect\citeauthoryear{Emmanoulopoulos et 
al.}{2011}]{2011MNRAS.415.1895E} Emmanoulopoulos D., Papadakis I.~E., 
McHardy I.~M., Nicastro F., Bianchi S., Ar{\'e}valo P., 2011, MNRAS, 415, 
1895
\bibitem[\protect\citeauthoryear{Fabian et al.}{2002}]{2002MNRAS.331L..35F} 
Fabian A.~C., Ballantyne D.~R., Merloni A., Vaughan S., Iwasawa K., Boller 
T., 2002, MNRAS, 331, L35
\bibitem[\protect\citeauthoryear{Fabian et al.}{2009}]{2009Natur.459..540F} 
Fabian A.~C., et al., 2009, Natur, 459, 540
\bibitem[\protect\citeauthoryear{Fabian et al.}{2012}]{2012MNRAS.419..116F} 
Fabian A.~C., et al., 2012, MNRAS, 419, 116
\bibitem[\protect\citeauthoryear{Fink et 
al.}{1997}]{1997A&A...317...25F} Fink H.~H., Walter R., Schartel N., Engels D., 1997, A\&A, 317, 25
\bibitem[\protect\citeauthoryear{George 
\& Fabian}{1991}]{1991MNRAS.249..352G} George I.~M., Fabian A.~C., 1991, MNRAS, 249, 352
\bibitem[\protect\citeauthoryear{Gierli{\'n}ski 
\& Done}{2004}]{2004MNRAS.349L...7G} Gierli{\'n}ski M., Done C., 2004, MNRAS, 349, L7
\bibitem[\protect\citeauthoryear{Grupe et al.}{2004}]{2004AJ....127..156G} 
Grupe D., Wills B.~J., Leighly K.~M., Meusinger H., 2004, AJ, 127, 156
\bibitem[\protect\citeauthoryear{Haardt 
\& Maraschi}{1993}]{1993ApJ...413..507H} Haardt F., Maraschi L., 1993, ApJ, 413, 507
\bibitem[\protect\citeauthoryear{Hwang 
\& Bowyer}{1997}]{1997ApJ...475..552H} Hwang C.-Y., Bowyer S., 1997, ApJ, 475, 552
\bibitem[\protect\citeauthoryear{Kalberla et 
al.}{2005}]{2005A&A...440..775K} Kalberla P.~M.~W., Burton W.~B., Hartmann D., Arnal E.~M., Bajaja E., Morras R., P{\"o}ppel W.~G.~L., 2005, A\&A, 440, 775 
\bibitem[\protect\citeauthoryear{King, Pringle, 
\& Hofmann}{2008}]{2008MNRAS.385.1621K} King A.~R., Pringle J.~E., Hofmann J.~A., 2008, MNRAS, 385, 1621
\bibitem[\protect\citeauthoryear{Kokubun et 
al.}{2007}]{2007PASJ...59S..53K} Kokubun M., et al., 2007, PASJ, 59, 53
\bibitem[\protect\citeauthoryear{Laor}{1991}]{1991ApJ...376...90L} Laor A., 
1991, ApJ, 376, 90
\bibitem[\protect\citeauthoryear{Larsson et 
al.}{2008}]{2008MNRAS.384.1316L} Larsson J., Miniutti G., Fabian A.~C., 
Miller J.~M., Reynolds C.~S., Ponti G., 2008, MNRAS, 384, 1316
\bibitem[\protect\citeauthoryear{Lawrence}{2011}]{2011arXiv1110.0854L} 
Lawrence A., 2011, arXiv, arXiv:1110.0854 
\bibitem[\protect\citeauthoryear{Lawrence 
\& Papadakis}{1993}]{1993ApJ...414L..85L} Lawrence A., Papadakis I., 1993, ApJ, 414, L85
\bibitem[\protect\citeauthoryear{Lawrence 
\& Elvis}{2010}]{2010ApJ...714..561L} Lawrence A., Elvis M., 2010, ApJ, 714, 561
\bibitem[\protect\citeauthoryear{Lynden-Bell}{1969}]{1969Natur.223..690L} 
Lynden-Bell D., 1969, Natur, 223, 690
\bibitem[\protect\citeauthoryear{Middleton, Done, 
\& Gierli{\'n}ski}{2007}]{2007MNRAS.381.1426M} Middleton M., Done C., Gierli{\'n}ski M., 2007, MNRAS, 381, 1426
\bibitem[\protect\citeauthoryear{Miniutti 
\& Fabian}{2004}]{2004MNRAS.349.1435M} Miniutti G., Fabian A.~C., 2004, MNRAS, 349, 1435
\bibitem[\protect\citeauthoryear{Nandra et al.}{2007}]{2007MNRAS.382..194N} 
Nandra K., O'Neill P.~M., George I.~M., Reeves J.~N., 2007, MNRAS, 382, 194
\bibitem[\protect\citeauthoryear{Nardini et 
al.}{2011}]{2011MNRAS.410.1251N} Nardini E., Fabian A.~C., Reis R.~C., 
Walton D.~J., 2011, MNRAS, 410, 1251
\bibitem[\protect\citeauthoryear{Osterbrock 
\& Pogge}{1985}]{1985ApJ...297..166O} Osterbrock D.~E., Pogge R.~W., 1985, ApJ, 297, 166
\bibitem[\protect\citeauthoryear{Ponti et al.}{2011}]{2011arXiv1112.2744P} 
Ponti G., Papadakis I., Bianchi S., Guainazzi M., Matt G., Uttley P., 
Bonilla F., Nuria, 2011, arXiv, arXiv:1112.2744
\bibitem[\protect\citeauthoryear{Porquet et 
al.}{2004}]{2004A&A...422...85P} Porquet D., Reeves J.~N., O'Brien P., Brinkmann W., 2004, A\&A, 422, 85
\bibitem[\protect\citeauthoryear{Pounds, Done, 
\& Osborne}{1995}]{1995MNRAS.277L...5P} Pounds K.~A., Done C., Osborne J.~P., 1995, MNRAS, 277, L5
\bibitem[\protect\citeauthoryear{Rees}{1984}]{1984ARA&A..22..471R} Rees M.~J., 1984, ARA\&A, 22, 471
\bibitem[\protect\citeauthoryear{Remillard 
\& McClintock}{2006}]{2006ARA&A..44...49R} Remillard R.~A., McClintock J.~E., 2006, ARA\&A, 44, 49
\bibitem[\protect\citeauthoryear{Reynolds 
\& Fabian}{2008}]{2008ApJ...675.1048R} Reynolds C.~S., Fabian A.~C., 2008, ApJ, 675, 1048 
\bibitem[\protect\citeauthoryear{Risaliti et 
al.}{2009}]{2009ApJ...705L...1R} Risaliti G., et al., 2009, ApJ, 705, L1
\bibitem[\protect\citeauthoryear{Romano et al.}{2002}]{2002ApJ...564..162R} 
Romano P., Turner T.~J., Mathur S., George I.~M., 2002, ApJ, 564, 162
\bibitem[\protect\citeauthoryear{Ross 
\& Fabian}{2005}]{2005MNRAS.358..211R} Ross R.~R., Fabian A.~C., 2005, MNRAS, 358, 211
\bibitem[\protect\citeauthoryear{R{\'o}{\.z}a{\'n}ska et 
al.}{2004}]{2004ApJ...600...96R} R{\'o}{\.z}a{\'n}ska A., Czerny B., 
Siemiginowska A., Dumont A.-M., Kawaguchi T., 2004, ApJ, 600, 96
\bibitem[\protect\citeauthoryear{Satyapal et 
al.}{2005}]{2005ApJ...633...86S} Satyapal S., Dudik R.~P., O'Halloran B., 
Gliozzi M., 2005, ApJ, 633, 86
\bibitem[\protect\citeauthoryear{Schmoll et 
al.}{2009}]{2009ApJ...703.2171S} Schmoll S., et al., 2009, ApJ, 703, 2171
\bibitem[\protect\citeauthoryear{Schurch, Done, 
\& Proga}{2009}]{2009ApJ...694....1S} Schurch N.~J., Done C., Proga D., 2009, ApJ, 694, 1
\bibitem[\protect\citeauthoryear{Scott et al.}{2011}]{2011MNRAS.417..992S} 
Scott A.~E., Stewart G.~C., Mateos S., Alexander D.~M., Hutton S., Ward M.~J., 2011, MNRAS, 417, 992
\bibitem[\protect\citeauthoryear{Shakura 
\& Sunyaev}{1973}]{1973A&A....24..337S} Shakura N.~I., Sunyaev R.~A., 1973, A\&A, 24, 337
\bibitem[\protect\citeauthoryear{Takahashi, Hayashida, 
\& Anabuki}{2010}]{2010PASJ...62.1483T} Takahashi H., Hayashida K., Anabuki N., 2010, PASJ, 62, 1483
\bibitem[\protect\citeauthoryear{Turner, George, 
\& Nandra}{1998}]{1998ApJ...508..648T} Turner T.~J., George I.~M., Nandra K., 1998, ApJ, 508, 648
\bibitem[\protect\citeauthoryear{Turner et al.}{2001}]{2001ApJ...548L..13T} 
Turner T.~J., et al., 2001, ApJ, 548, L13
\bibitem[\protect\citeauthoryear{Turner et al.}{2002}]{2002ApJ...568..120T} 
Turner T.~J., et al., 2002, ApJ, 568, 120
\bibitem[\protect\citeauthoryear{Vaughan 
\& Fabian}{2004}]{2004MNRAS.348.1415V} Vaughan S., Fabian A.~C., 2004, MNRAS, 348, 1415 
\bibitem[\protect\citeauthoryear{Vaughan et 
al.}{2002}]{2002MNRAS.337..247V} Vaughan S., Boller T., Fabian A.~C., 
Ballantyne D.~R., Brandt W.~N., Tr{\"u}mper J., 2002, MNRAS, 337, 247
\bibitem[\protect\citeauthoryear{Vaughan et 
al.}{2003}]{2003MNRAS.345.1271V} Vaughan S., Edelson R., Warwick R.~S., 
Uttley P., 2003, MNRAS, 345, 1271 
\bibitem[\protect\citeauthoryear{Vennes et al.}{1995}]{1995ApJ...448L...9V} 
Vennes S., Polomski E., Bowyer S., Thorstensen J.~R., 1995, ApJ, 448, L9
\bibitem[\protect\citeauthoryear{Walton, Reis, 
\& Fabian}{2010}]{2010MNRAS.408..601W} Walton D.~J., Reis R.~C., Fabian A.~C., 2010, MNRAS, 408, 601
\bibitem[\protect\citeauthoryear{Walton et al.}{2012}]{2012arXiv1202.5193W} 
Walton D.~J., Reis R.~C., Cackett E.~M., Fabian A.~C., Miller J.~M., 2012, 
arXiv, arXiv:1202.5193
\bibitem[\protect\citeauthoryear{Wilms, Allen, 
\& McCray}{2000}]{2000ApJ...542..914W} Wilms J., Allen A., McCray R., 2000, ApJ, 542, 914
\bibitem[\protect\citeauthoryear{Wilkins 
\& Fabian}{2011}]{2011MNRAS.414.1269W} Wilkins D.~R., Fabian A.~C., 2011, MNRAS, 414, 1269
\bibitem[\protect\citeauthoryear{Wisotzki et 
al.}{1995}]{1995A&A...297L..55W} Wisotzki L., Dreizler S., Engels D., Fink H.-H., Heber U., 1995, A\&A, 297, L55
\bibitem[\protect\citeauthoryear{Zoghbi, Fabian, 
\& Gallo}{2008}]{2008MNRAS.391.2003Z} Zoghbi A., Fabian A.~C., Gallo L.~C., 2008, MNRAS, 391, 2003
\bibitem[\protect\citeauthoryear{Zoghbi et al.}{2010}]{2010MNRAS.401.2419Z}
Zoghbi A., Fabian A.~C., Uttley P., Miniutti G., Gallo L.~C., Reynolds
C.~S., Miller J.~M., Ponti G., 2010, MNRAS, 401, 2419
\end{thebibliography}
\end{document}